\documentclass[twocolumn,showpacs,amsfonts]{revtex4-1}

\usepackage{amsmath}
\usepackage{latexsym}
\usepackage{float}
 \usepackage{amssymb}
\usepackage{graphicx}
\usepackage{textcomp}
\usepackage{color}
\usepackage{hyperref}
\usepackage{subfigure}

\def\ba{\begin{eqnarray}}
\def\ea{\end{eqnarray}}
\def\be{\begin{equation}}
\def\ee{\end{equation}}
\def\bm{\begin{math}}
\def\me{\end{math}}

\newcommand{\dummy}

\begin{document}

\title{Ballistic Aggregation in Systems of Inelastic Particles: 
Cluster growth, structure and aging}
\author{ Subhajit Paul and Subir K. Das$^{*}$} 
\affiliation{Theoretical Sciences Unit, Jawaharlal Nehru Centre for Advanced Scientific Research,
 Jakkur P.O, Bangalore 560064, India}

\date{\today}

\begin{abstract}
~ We study far-from-equilibrium dynamics in models of freely cooling granular gas and 
ballistically aggregating compact clusters.
For both the cases, from event driven molecular dynamics simulations we have presented detailed 
results on structure and dynamics in space dimensions $d=1$ and $2$. 
Via appropriate 
analyses it has been confirmed that the ballistic aggregation mechanism applies in $d=1$ granular gases as well. 
Aging phenomena for this 
mechanism, in both the dimensions, have been studied via the two-time density 
autocorrelation function. This quantity is demonstrated 
to exhibit scaling property similar to that in the standard phase transition kinetics. 
The corresponding functional forms have been 
quantified and discussed in connection with the structural properties. Our results on aging 
establish a more complete equivalence between the granular gas 
and the ballistic aggregation models in $d=1$.
\end{abstract}


\maketitle

\section{Introduction}
~Structure and dynamics during cooling in systems of inelastically colliding particles have been of much research interest 
\cite{ara,bri,gold,brito,lud1,ben,nie,das1,das2,sin1,sin2,bod,paul,trizac1,trizac2,trizac3,lipow,haff,herbst,ulrich,car,hummel1}. 
An importance of this topic stems 
from the relevance of it in the agglomeration of cosmic dust \cite{car}. 
Two models in this context have been of significant importance, viz., 
the granular gas model (GGM) 
and ballistic aggregation model (BAM).
In the BAM, following a collision between two freely moving clusters, the colliding partners 
form a single larger object. In one dimension this corresponds to the sticky gas.
While collisions trigger clustering immediately in the case of BAM, for the GGM (with coefficient
of restitution $0<e<1$) 
the system remains in a homogeneous density state during an initial period, 
referred to as the homogeneous cooling state (HCS) \cite{gold,haff}. 
The dynamics in the latter then 
crosses over to an inhomogeneous cooling state (ICS) \cite{gold}, where particle-poor and particle-rich domains emerge.
Time scale for such a crossover gets shorter with the decrease of $e$.
These domains or clusters may grow for indefinite
period of time if the system size is thermodynamically large \cite{hummel1}. 
Thus, even if not a phase transition,
it is quite natural to study clustering phenomena in these models
from the perspectives of phase transition kinetics \cite{bray,onuki,puri}. 
\par
~~In problems of phase transitions \cite{bray}, having been quenched from a homogeneous state to a 
state inside the miscibility gap, as a system proceeds towards the new
equilibrium, one is interested in understanding the domain pattern \cite{bray}, its growth \cite{bray} and aging \cite{puri}. 
Typically, a pattern is characterized via the two-point equal-time correlation function \cite{bray}
$C$, which, in an isotropic situation, is calculated as \cite{bray}
($r$ being the scalar distance between two points)
\begin{equation}\label{eq1}
 C(r,t) = \langle\psi(\vec{r},t) \psi(\vec{0},t)\rangle - \langle\psi(\vec{r},t)\rangle\langle\psi(\vec{0},t)\rangle,
\end{equation}
where $\psi$ is a space ($\vec r$) and time ($t$) dependent order parameter. For a vapor-liquid transition, which granular
systems have resemblance with, $\psi$ is related to the local density.
For a self-similar pattern,  $C(r,t)$ and its Fourier transform, $S(k,t)$ ($k$ being the magnitude of the wave vector),
the structure factor, obey the scaling properties \cite{bray}
\begin{equation}\label{correl_sf}
 C(r,t) \equiv \tilde{C} (r/\ell),~S(k,t) \equiv \ell ^{d} \tilde{S}(k\ell),
\end{equation}
where $\tilde{C}$ and $\tilde{S}$ are time-independent master functions \cite{bray}.
These dynamic scalings reflect the fact that structures at two
different times are similar, apart from a change in length scale
$\ell$, the average size of domains, that
grows with time as \cite{bray}
\begin{equation}\label{glaw}
\ell\sim t^{\alpha}.
\end{equation}
\par
~~For the study of aging property, one considers a two-time
autocorrelation function \cite{fish,maz}
\begin{equation}\label{eq5}
C_{\rm ag}(t,t_w) = \langle\psi(\vec{r},t) \psi(\vec{r},t_{w})\rangle - \langle\psi(\vec{r},t)\rangle\langle\psi(\vec{r},t_{w})\rangle,
\end{equation}
where $t_{w}$ ($\leq t$) is referred to as the waiting time or age of the system \cite{puri}. Unlike the equilibrium situation,
the decay of $C_{\rm ag}(t,t_{w})$ gets slower with the increase of $t_w$, when plotted vs $t-t_w$, 
since there is no time translation invariance in an evolving system.
In kinetics of phase transitions,
 $C_{\rm ag}(t,t_{w})$
follows a scaling relation \cite{puri,fish}
\begin{equation}\label{aging1}
C_{\rm ag}(t,t_{w}) \equiv \tilde{C}_{\rm ag}(\ell/\ell_{w}),
\end{equation}
where $\ell_{w}$ is the domain size at $t_{w}$ and $\tilde{C}_{\rm ag}$ is a master function \cite{puri}
that typically exhibits power-law decay as a function of $\ell/\ell_{w}$.
Examination of these facts for systems as nontrivial as those consisting of inelastic particles
should be of genuine interest, to gain an universal picture of the concepts of nonequilibrium statistical mechanics,
since these systems continuously dissipate kinetic energy.

Like in phase transitions, in the case of inelastic particles also, 
considered to be hard spheres in many theoretical studies,
power-law growths have been observed \cite{lud1,das1,das2,paul,sin1,sin2}. 
In phase transitions, $\ell$ can be connected to the interfacial energy.
Even though a connection with interfacial energy does not exist here,
for the BAM the average cluster mass ($m$) has been related with
the average kinetic energy ($E$) \cite{car}.
In many problems of coarsening, the growth exponent $\alpha$ depends upon the transport
mechanism and system dimensions \cite{bray}. In the case of BAM also
time-dependence of
$m$ has been predicted to have strong influence from dimension \cite{car}. Despite these advancements,
many questions remain open, including the issue related to the equivalence between BAM and GGM.
\par
~~For the growth of  
$m$ ($\sim \ell^d$, $d$ being the system dimension) via the ballistic aggregation mechanism,
a scaling theory predicts \cite{car}
\begin{equation}\label{mass_t}
m \sim \frac{1}{E} \sim t^{\frac{2d}{d+2}}.
\end{equation}
In $d=1$, this has been confirmed via simulations, for both BAM and GGM cases \cite{ben,nie,sin1,sin2,car}.
In higher 
dimensions, on the other hand, the status is not satisfactory with respect to
the equivalence between the two models. 
Even though the time-dependence of $E$ is reported to be consistent with Eq. (\ref{mass_t}), 
for the growth of $m$ in GGM, there exists evidence for dimension independence \cite{paul}. 
This raises question whether the complete validity of
Eq. (\ref{mass_t}) in $d=1$, for both BAM and GGM cases, is accidental. 
Thus, even in this dimension, direct confirmation of the mechanism for GGM is essential, to
draw a conclusion on the equivalence \cite{ben} between the two models.
For $d>1$, strictly speaking, even for BAM, 
the time dependence in Eq. (\ref{mass_t}) requires modification, since in that case one expects
isolated fractal clusters \cite{mid1} with fractal dimension $d_f$ ($<d$) such that $m\sim \ell^{d_f}$.
This fact is not included in Eq. (\ref{mass_t}) and
due to technical difficulties, in existing simulation studies also 
spherical (compact) structural assumption \cite{trizac1,trizac2} 
of the growing clusters became necessary.
\par
~~Furthermore, while some aspects of kinetics have been studied, 
aging property \cite{puri,fish,maz,yeu,henkel,cor,ahm,maj1} of the density field and its
connection to pattern and growth did not receive attention for these models, though important \cite{hummel2}.
To the best of our knowledge,
there exists only one study \cite{balda} that addresses scaling property
of the two-time correlation function in the granular-matter context. 
This, however, considers aging in a different quantity, for a
model different from the one considered here.
\par
~~ Here note that various scaling properties that have been established with respect to aging are related to
approach of a far-from-equilibrium system to an equilibrium state, like in phase transitions.
Given that systems of inelastically colliding particles are always out of equilibrium, examination
of the validity of these properties in such systems should be of fundamental importance. If scaling exists, it is
of interest then to compare the scaling functions associated with GGM and BAM cases,
to establish a more complete equivalence. 
\par
~~ In this work, our primary objective is to identify the scaling property related to aging in ballistic aggregation.
For this purpose, 
it has been shown, via a state-of-the-art
dynamic renormalization group theoretical method of analysis \cite{rol},
that the growth law for one-dimensional GGM in ICS is same as that for the BAM.
In this dimension, we also directly show that the mechanism of aggregation in GGM is indeed ballistic. 
These, along with our results on aging, establish a more complete equivalence
between GGM and BAM in $d=1$.
On the other hand, we have pointed out vast differences between the structure and dynamics of GGM
and the corresponding theoretical expectations for $d=2$ BAM. For the latter dimension, thus, for aging property 
we work only with 
the BAM. 
We show that, irrespective of the dimension,  the above scaling property of the autocorrelation 
function holds for both the models as long as the growth occurs via ballistic aggregation.
These results are discussed with reference to the picture in standard phase transitions.
The functional forms of $\tilde{C}_{\rm ag}$ have been estimated and understood via 
analyses of the structure \cite{yeu}. 
In $d=2$, via accurate analyses, we also check the validity of a hyperscaling relation involving
the decay of energy and growth of clusters for the BAM.
\par
~~ The rest of the paper is organised in the following way. We discuss the model and methods in Section II. The results are presented in 
Section III. Finally, Section IV concludes the paper with a summary and outlook.

\section{Model and Methods}
 ~~ For the GGM we use the following update rule for (hard) particle velocities.
The post and pre-collisional velocities of the particles are related via \cite{gold,all,rap}
\begin{equation}\label{vel_chngi}
 \vec{v}^{~\prime}_{i} = \vec{v}_{i} -\Big(\frac{1+e}{2}\Big)[\hat{n}\cdot(\vec{v}_{i}-\vec{v}_{j})]\hat{n},
\end{equation}
\begin{equation}\label{vel_chngj}
 \vec{v}^{~\prime}_{j} = \vec{v}_{j} -\Big(\frac{1+e}{2}\Big)[\hat{n}\cdot(\vec{v}_{j}-\vec{v}_{i})]\hat{n},
\end{equation}
where ($^\prime$) stands for the post event, $\vec{v}_{i}$ and $\vec{v}_{j}$ are velocities of particles $i$ and $j$, 
respectively, and $\hat{n}$ is the unit vector
in the direction of the relative position of the particles $i$ and $j$. 
With this model, we perform event driven \cite{all,rap} molecular dynamics simulations where an event is a collision. 
In this method, between two collisions, since there are no inter-particle 
interaction or external potential, particles move with constant velocities 
till the next collision, which is appropriately identified after every event. 
\par
~~For the 
BAM case \cite{car}, following every collision, mass of the product particle 
increases, which was appropriately incorporated in the collision rule. 
For the BAM in $d=2$ we use the same circular approximation of the product clusters 
as in the previous studies \cite{pathak}, which will be briefly discussed later.
Typically, in such event driven simulations, time is specified in two different ways, viz., 
by using the number of collisions per particle ($\tau$) and actual time ($t$), 
the latter being calculated by keeping track of the free time between successive collisions. 
In this work, we will use the latter.
\par
~~ A serious problem faced in event driven simulations of the GGM is the inelastic collapse \cite{mcn}. 
This phenomenon is related to the fact that for very low values of the
relative velocity collisions keep happening only among a small group of neighboring particles, 
thereby essentially providing no progress in time. The problem 
is more severe in lower dimension, since fewer particles are needed to satisfy 
the corresponding condition. There can be two ways to avoid such singularity in collision numbers, 
viz., setting the value of $e$, for the collision partners with relative velocities 
less than a threshold value $\delta$, to either $0$ or $1$. We adopt the 
latter \cite{ben,mcn,camp,lud2} given that in the experimental situation value of $e$
increases with the decrease of the relative velocity \cite{sin1,sin2,bod,ram}. In $d=2$, however,
we set $\delta$ to zero, since the problem is less severe in higher dimension and so,
significantly large cluster sizes can be accessed without encountering such events.
\par
~~ All our results will be presented from simulations with periodic boundary 
conditions and density of particles starting with 
$\rho=N/L^d=0.30$, $N$ being the number of particles and $L$ the 
linear dimension of the system, except for the $d=2$ GGM for which we choose $\rho=0.37$. 
Given that the the particles have diameter unity (to start with), in $d=2$ these numbers for
particle density correspond to packing fractions $0.235$ (BAM) and $0.29$ (GGM).
\par
~~In the case of GGM, clusters were appropriately identified as regions with density 
above a critical value $\rho_{c}$ ($=0.5$). Higher values of $\rho_c$
also provide similar results, deviating from each other only by a multiplicative factor.
The end to end distance for a cluster along any direction provides
a cluster length ($\ell_{c}$). In $d=1$, the number of particles within these boundaries is the mass ($m_c$)
of that cluster. In $d=2$, one needs to appropriately identify the closed boundaries of the clusters,
to calculate the mass.
For the BAM case, information on the cluster length and mass 
are contained in the particle radius.
The average values of the above mentioned quantities were obtained from the first 
moments of the corresponding distributions. Ideally, $\ell$ should equal $m^{1/d}$, 
but in the case of GGM it takes time for a cluster to settle down to a 
particular density value. Thus, equality holds only at late time. 
For the calculation of the correlation functions and structure factors \cite{das1,das2}, 
the order-parameter $\psi$ at a point was assigned a value $+1$ if the density (calculated
by counting the number of nearest neighbors) there was higher 
than $\rho_{c}$, else $-1$. The average length can be calculated from the 
scaling property of $C(r,t)$ or $S(k,t)$ as well.

\section{Results}
~~We divide the section into sub-parts {\textbf{A}} and {\textbf{B}}. In subsection  {\textbf{A}} we present results 
from $d=1$ and $d=2$ results are shown in subsection {\textbf{B}}.
\subsection{$d=1$}
~~As mentioned above, in this dimension, via accurate analyses, we first confirm an equivalence between the BAM
and GGM, with respect to the energy decay, growth law and mechanism. These results are followed by those 
for aging property. As we will see, the latter, in addition to being of separate importance, will make the above mentioned 
equivalence more complete.

\begin{figure}[htb]
\centering
\includegraphics*[width=0.485\textwidth]{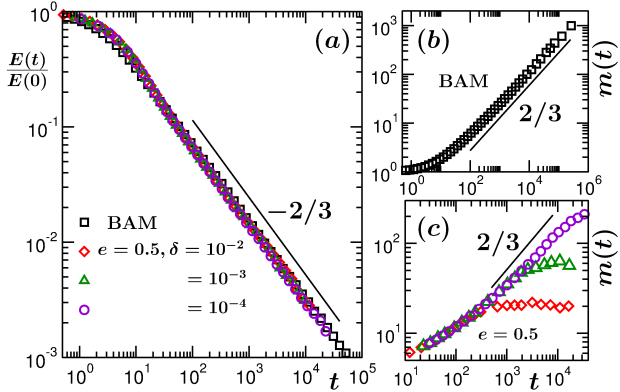}
\caption{\label{fig1} (a) Plots of energy decay as a function of time, for BAM and GGM cases. 
For the GGM case, value of $e$ has been fixed to $0.5$ and results for several choices of $\delta$
are presented. (b) Average cluster mass, $m$, is plotted vs time, for BAM.
(c) Same as (b) but for GGM with multiple values of $\delta$ as in (a). The solid lines in these figures 
correspond to various power-laws, exponents for which
are mentioned. The starting number of particles for BAM and GGM are respectively $160000$ and $10000$. 
Rest of the simulations for GGM are done with $20000$ particles and
$\delta=5\times10^{-5}$. All results are for $d=1$.}
\end{figure}

~~In Figure \ref{fig1}(a) we show the decay of energy, for both GGM and BAM, 
as a function of time, on a log-log scale. The BAM results, 
for energy and mass (see Figure \ref{fig1}(b)), are
already understood. However, we present these for the sake of completeness, as well as to
facilitate the discussion that follows.
For the GGM, results for a few 
different cut-off values of the relative velocity are shown. For this case, in this dimension, all our 
results correspond to $e=0.5$. After a minor disagreement over 
brief initial period (corresponding to HCS in the GGM), all the results are consistent with each other, 
exhibiting power-law behavior over several decades in time, with the expected exponent $-2/3$. 
\par
~~In Figure \ref{fig1}(b) we plot $m$ as a function of $t$, on a log-log scale, for the 
BAM case. This shows a power-law growth with exponent $2/3$, validating Eq. (\ref{mass_t}). 
The $m$ vs $t$ results for the GGM are shown in Figure \ref{fig1}(c),
for the same values of $\delta$ as in Figure \ref{fig1}(a). An interesting observation here is 
that, for the GGM,  even though the energy decay follows $t^{-2/3}$ behavior till late 
for all values of $\delta$, the picture is different for the growth of mass. The growth stops earlier for larger
value of $\delta$, even though energy decay continues with the predicted functional form. 
This should not be a finite-size effect, since the saturation is $\delta$ dependent.
Rather, this has connection with late time declusterization phenomena \cite{sin1,sin2} that has been observed 
for relative velocity dependent $e$. 
Furthermore, the $m$ vs $t$ data, particularly for larger values of $\delta$, do not appear consistent with 
the exponent $2/3$. This discrepancy  
can possibly be due to the presence of substantial length at the beginning of the scaling regime. In such a situation,
confirmation of an exponent from a log-log plot requires data over several
decades in time \cite{maj2}. 
In absence of that, alternative accurate method of analysis is needed to obtain correct value of the exponent
\cite{rol,maj2,huse}. 
In any case, the observations above, with respect to the saturation of $m$,
further justify the need for direct identification
of the growth mechanism. Before moving to that we will accurately quantify the growth law. For this purpose,
in the following we will work with the length rather than the mass, since for the aging property we 
will need this latter quantity. Unless otherwise mentioned, in this subsection, all our results for the GGM,
 from here on, will be presented  for $\delta=5\times10^{-5}$.

\begin{figure}[htb]
\centering
\includegraphics*[width=0.49\textwidth]{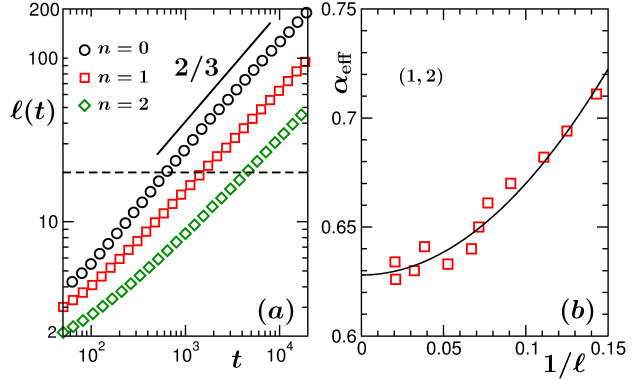}
\caption{\label{fig2} (a) Plots of $\ell$ vs $t$ for three different stages of renormalization.
The dashed horizontal line is for the extraction of times for the same length
at different levels of renormalization. The solid line represents a power-law with exponent $2/3$. 
(b) Plot of the effective exponent, obtained via the renormalization-group analysis using 
the combination $n=1$ and $2$,
vs the inverse of the original length. The solid line is a quadratic fit to the
simulation data.
All results correspond to the GGM in $d=1$.}
\end{figure}

\par
~~ We use a renormalization-group method of analysis \cite{rol} for the accurate quantification of the 
growth for the GGM. 
We consider a Kadanoff type block transformation \cite{golden} of the order
parameter. For this purpose, as mentioned in the context of calculation of the correlation functions,
we have mapped the density field to $\psi=\pm 1$. The blocking exercise then
becomes similar to that for the Ising model \cite{bray}.
At successive iterations of the transformation, order parameter over a length of $b$ particle diameters 
is averaged over and represented by a single point,
reducing the system size by a factor $b$, for which we choose the value $2$. 
Thus, a particular value of $\ell$ in different levels ($n$) of renormalization will be 
obtained at different times, viz., for $n=p$ and $p+1$ one writes \cite{rol}
\begin{equation}\label{rgl}
 \ell(p,t)= \ell (p+1,b^{1/\alpha}t).
\end{equation}
This is demonstrated in Figure \ref{fig2}(a), where, in addition to the original data ($n=0$),
we have presented length vs time plots for renormalizations with $n=1$ and $2$. 
The horizontal line in this figure is
related to the estimation of times for the same length scale
for different values of $n$.
From the shifting or
scaling of time, due to the scaling in length, 
the growth exponent can be estimated. However, because of technical reasons the true exponent will be realized
only in the limit $\ell\rightarrow\infty$ and for finite time we will denote it by $\alpha_{\rm eff}$.

\begin{figure}[htb]
\centering
\includegraphics*[width=0.485\textwidth]{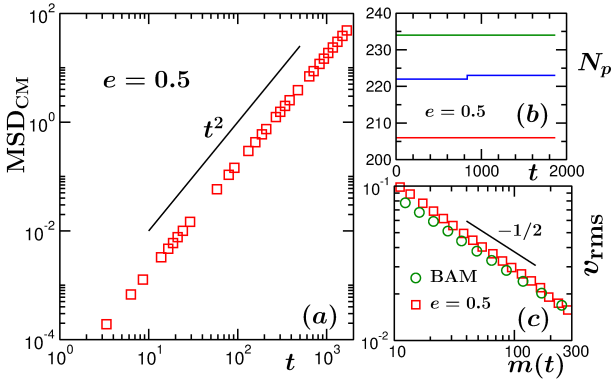}
\caption{\label{fig3} (a) Mean-squared-displacement of the centre of mass of a typical cluster, for GGM, is plotted vs
time, on a log-log scale. The solid line corresponds to ballistic
motion. (b) Number of particles in a few different clusters, for GGM, are plotted vs translated time, before 
they undergo collisions.
(c) Root-mean-squared velocity of the clusters
are plotted vs $m$, for both BAM and GGM. 
The solid line is a power-law decay, exponent being mentioned.
All results are from $d=1$.}
\end{figure}

Estimated values of $\alpha_{\rm eff}$, from the combination involving $n=1$ and $2$,
are presented in Figure \ref{fig2}(b), vs $1/\ell$, which indeed have time
dependence \cite{das3}.
The time dependence is due to the nonscaling early time transient and presence of a large off-set when
scaling is reached. When such time dependence exits, as already stated,
$\alpha$ should be estimated from the convergence of the  data in the $\ell \rightarrow \infty$
limit. By looking at the trend of the data set, we have fitted it to the form 
$\alpha_{\rm eff}=\alpha+a/\ell^2$, that provides convergence to
$\alpha\simeq 0.63$. This is very close to
the ballistic aggregation \cite{car} value $2/3$. 
To check, whether this minor deviation of the simulation data from the theoretical expectation is
a true fact, one needs to study other values of $e$ as well. Such a systematic study we leave out
for a future work. The deviation could be due to the finite-size effects and $\delta$-dependent saturation.
\par
~~ Having identified the growth exponent for the GGM, we, in Figure \ref{fig3}, 
identify the mechanism. The growth exponent $2/3$ can be obtained from a (nonequilibrium)
kinetic theory for ballistic aggregation \cite{car,mid1,pat}. 
As the name suggests, the growth occurs in this mechanism due to collisions
among clusters and between collisions the clusters move with constant velocities.
Since the particles in our models are noninteracting, 
it is understandable that the clusters in the BAM will move ballistically between collisions. 
In the GGM case also,
following more and more collisions, particles within a cluster may move parallel
to each other, providing collective directed motion. However, when a cluster moves through a vapor region,
growth in this case may occur due to random deposition of particles on them. It is
then necessary to check if at late enough time the motion of the clusters, during the interval
between two big mass enhancing collisions, are ballistic and during that period
the growth of the clusters is negligible.
\par
~~ In part (a) of Figure \ref{fig3} we show the mean-squared-displacement of the centre of mass (CM), ${\rm MSD_{CM}}$,
of a cluster, calculated as \cite{han}
\begin{equation}\label{msd_cm}
\mbox{MSD}_{\mbox{CM}}=\langle |{\vec R}_{\mbox{CM}}(t)-{\vec R}_{\mbox{CM}}(0)|^2 \rangle,
\end{equation}
${\vec R}_{\mbox{CM}}$ being the time-dependent location of the CM
of the cluster, for GGM, over an extended period of time, before it undergoes a collision with 
another cluster. On the log-log scale, a very robust $t^2$ behaviour is visible,
confirming ballistic motion \cite{han}. In Figure \ref{fig3}(b) we show number of particles in a 
few clusters, as a function of translated time. The constant values over long time confirm that
the mechanism of growth in the GGM is indeed ballistic aggregation. 
\par
~~ The mass part of Eq. (\ref{mass_t}) can be derived from \cite{car,mid1,pat}
\begin{equation}\label{kinetic}
 \frac{dn_{c}}{dt} = -\ell^{d-1} v_{\rm rms} n_{c}^2,
\end{equation}
where $n_{c}$ is the cluster density and $v_{\rm rms}$ is the root-mean-squared velocity of the clusters. 
An exponent $2/3$ requires $v_{\rm rms} \sim m^{-1/2}$, an outcome for uncorrelated cluster motion \cite{pat},
can be realized for Boltzmann distribution of cluster kinetic energies \cite{mid1,trizac2}. In Figure \ref{fig3}(c),
we plot $v_{\rm rms}$ vs $m$, for BAM as well as GGM, both of which show reasonable
consistency with the 
requirement. Here note that at low values of particle density, strong velocity correlation
is expected to appear, since the collisions are less random, 
leading to deviation from such Boltzmann distribution picture. At high density,
the collisions are random and such a picture is a good approximation. The reasonable validity of $v_{\rm rms}$
with the $m^{-1/2}$ form, that is observed,
should, however, be checked for other values of $e$ for the GGM to see if there exists complex density dependence.
Any deviation, though
does not invalidate the ballistic aggregation, can bring in change in the growth exponent. Here, as a passing remark,
we mention that for the ballistic aggregation of fractal clusters in $d$ dimensions, with $v_{\rm rms} \sim m^{-\gamma}$,
the exponent for the time dependence of mass will have the form
\begin{equation}\label{frac_exponent}
\zeta=\frac{d_f}{1-d+d_f(1+\gamma)},
\end{equation}
if Eq. (\ref{kinetic}) is a good starting point. Given that for the present problem $d_f=d=1$ and our estimate of $\gamma$ 
for the GGM is $0.55$, $\zeta=0.645$, in agreement with the conclusion from Figure \ref{fig2}(b). We obtain similar 
result for $\zeta$ via analysis of the instantaneous exponent. For $d=2$, we will adopt this method, instead of 
the renormalization group.

\begin{figure}[htb]
\centering
\includegraphics*[width=0.46\textwidth]{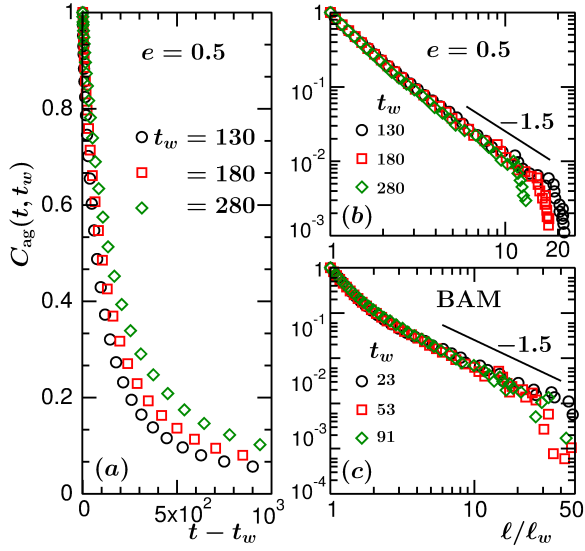}
\caption{\label{fig4} (a) Plots of the autocorrelation function, vs ($t-t_w$), for three different choices of $t_w$,
as mentioned, for the $d=1$ GGM.
(b) Log-log plots of $C_{\rm ag}(t,t_w)$ vs $\ell/\ell_w$, using the data sets in (a). 
(c) Same as (b) but for the $d=1$ BAM case. The values of $t_w$ are mentioned on the figure.
The solid lines in (b) and (c) represent power-law decays with exponent $\lambda=1.5$.}
\end{figure}

\par
~~Next we present results for the aging property \cite{puri,fish}. We stress again, not only in 
the granular matter context, to the best of our knowledge, aging has not been studied previously for ballistic
aggregation in any other system. In Figure \ref{fig4}(a) we plot $C_{\rm ag}(t,t_w)$ vs $t-t_{w}$, 
for a few different values of $t_{w}$, for the GGM. As expected, no time translation invariance is noticed
which is an equilibrium \cite{han} (or steady-state) property. Sticky gas (BAM) results are similar (not shown).
In Figure \ref{fig4}(b) we show $C_{\rm ag}(t,t_{w})$ vs $\ell/\ell_{w}$, on a log-log scale, for the GGM. 
Nice collapse of data from all chosen values of $t_w$ are seen, as in kinetics of 
phase transition. Deviations of the data sets from the master curve, appearing earlier for larger
values of $t_w$, are due to finite-size effects \cite{mid2}.
In phase transitions, the system moves towards an 
equilibrium state. Interestingly, similar scaling is observed 
in the present case, despite the fact that the system is continuously dissipating kinetic energy.
Corresponding plots for the BAM are shown in Figure \ref{fig4}(c). 
Again, very good quality collapse is observed. In both the cases, power-law decays \cite{fish}

\begin{equation}\label{aging_scale}
\tilde{C}_{\rm ag} \sim x^{-\lambda};~ x=\ell/\ell_{w},
\end{equation}
of the scaling 
function are observed for $x>>1$, the exponent value, mentioned on the figures, 
being same (or close to each other) in the two cases. 
This further confirms the equivalence between the BAM and the GGM.
\par
~~In phase transitions, there exists a lower bound \cite{fish,yeu} for the value of $\lambda$, viz.,

\begin{equation}\label{aging_bound}
\lambda \ge \frac{d+\beta}{2},
\end{equation}
where $\beta$ is the exponent for the small wave number power-law behaviour of the structure factor:
\begin{equation}\label{sf_beta}
S(k,t)\sim k^{\beta}.
\end{equation}
The bound in (\ref{aging_bound}) was derived by Yeung, Rao and Desai (YRD) \cite{yeu}.
For this purpose, starting from the structure factors at times $t$ and $t_w$, YRD obtained
\begin{equation}\label{yrd_b}
C_{\rm ag}(t,t_w) \leq \ell^{d/2}\int_0^{2\pi/\ell} dk k^{d-1} [S(k,t_w)\tilde{S}(k\ell)]^{1/2}.
\end{equation}
The bound follows when Eq. (\ref{sf_beta}), for the small $k$ behavior of $S(k,t_w)$, is used in the above expression.
To check whether $\lambda$ in the present case also obeys the bound (\ref{aging_bound}), we analyze the structure. 

\begin{figure}[htb]
\centering
\includegraphics*[width=0.46\textwidth]{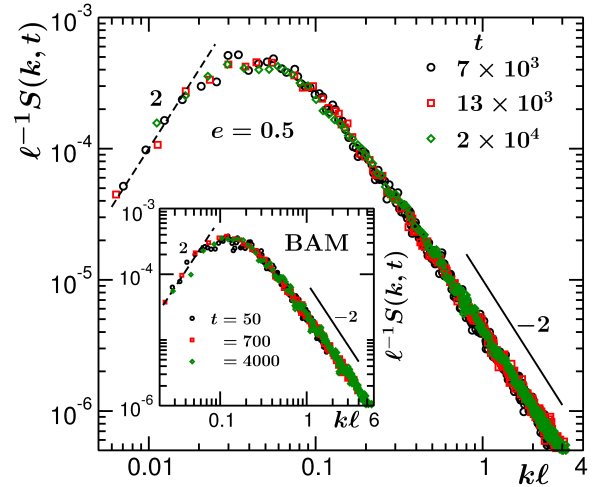}
\caption{\label{fig5} Scaling plot of the equal-time structure factors for the $d=1$ GGM. Here we have shown collapse of 
$S(k,t)/\ell(t)$, when plotted as a function of $y=k\ell(t)$, using data from three different times. 
Inset shows the same exercise as in the main frame but for $d=1$ BAM.
In the main frame as well as in the inset, the dashed and the solid lines correspond to 
$\sim y^2$ and $\sim y^{-2}$, respectively.}
\end{figure}

\par
~~ In Figure \ref{fig5} we show the scaling plot of the structure factors, viz., we plot $\ell^{-1}S(k,t)$ vs $k\ell$, 
for the GGM. Nice collapse of data from all different times imply 
structural self-similarity \cite{bray}.  The consistency of the long wave-vector data with $k^{-2}$ 
imply validity of the Porod law \cite{bray,porod,op1}
\begin{equation}\label{eq9}
S(k,t) \sim k^{-(d+1)},
\end{equation}
a consequence of short-distance singularity in $C(r,t)$, due to scattering
from sharp interfaces.
The small $k$ behaviour appears consistent with 
$\beta=2$. The behaviour for
the BAM structure factor, shown in the inset of Figure 5, is very similar. This value of $\beta$ was
predicted \cite{maja} for coarsening in Ising-like systems in $d=1$. The number is different for higher dimensions \cite{yeung2}.
The dimension dependent values of $\beta$ can be obtained \cite{Furu} from dynamical equation
of structure factor (starting from the Cahn-Hilliard equation \cite{Furu}) 
in $k$ space, by arguing that for $d=1$ thermal energy is dominant, whereas for
$d>1$ interfacial free energy takes over.
Agreement of our results with such
prediction is very interesting. The information on the consistency, for both short and long range
structures, between GGM and 
BAM, that these data sets convey, is further supportive of the presence of sharp interfaces,
compact clusters and ballistic aggregation in the GGM.
\par
~~ The observed value of $\beta$ sets the lower bound for $\lambda$ at $1.5$. Thus, this bound is obeyed in
both the cases and the actual values of the aging exponent in fact are very close to
this lower bound. Here note that
recently violation of such power-law decay of the autocorrelation function
was demonstrated \cite{ahm,maj1} for advective transport in fluid phase separations. Even for
conserved order parameter with diffusive dynamics, though power-law, the decays in $d>1$ are observed \cite{mid3} to be
significantly faster than the ones provided by the the (lower) bound (\ref{aging_bound}). However, in the latter example,
agreement with the bound gets better as the dimension decreases \cite{mid3}. 
With the lowering of
$d$, particularly for Ising kinetics, motion of the boundaries of domains (during no growth periods) gets restricted. 
However, since the mechanism is ballistic in the present problem, boundary movement does exist even during no growth period,
though decreases with the increase of mass, thus time. Nevertheless, the agreement with the lower bound is rather close.

\subsection{$d=2$}
~~In this subsection, first we briefly discuss the case of GGM, to convince ourselves that the growth in this case does not occur 
via the ballistic aggregation mechanism. Unlike the simulations of GGM in $d=1$, we do not use any nonzero cut-off value ($\delta$) 
for the relative velocity here. This is because, for high enough value of $e$, in this dimension, we are able to access relevant 
scaling regime without encountering an inelastic collapse \cite{paul}. 
\par
\begin{figure}[htb]
\centering
\includegraphics*[width=0.44\textwidth]{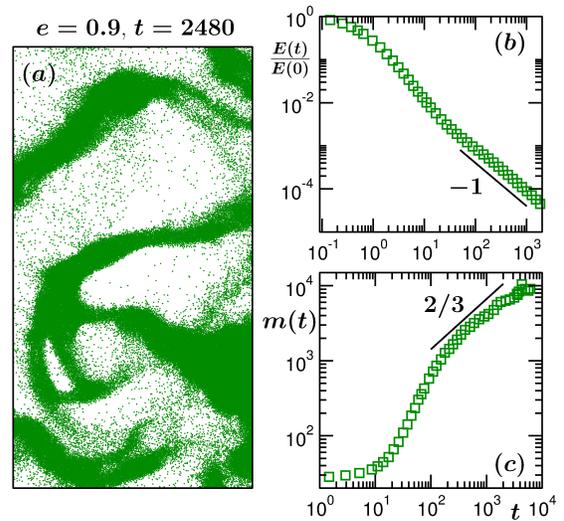}
\caption{\label{fig6} (a) A snapshot during the evolution of GGM with $e=0.9$ in $d=2$. Only a part of the snapshot is shown. 
The particles are marked by dots. (b) Log-log plot of the energy decay as a function of $t$, for the system in (a). (c) Log-log plot
 of mass vs time, for the GGM in (a). All results are for $L=512$. The solid lines in (b) and (c) represent power laws, 
exponents for which are mentioned.}
\end{figure}

~~In Figure \ref{fig6}(a) we show an evolution snapshot for the $d=2$ GGM with $e=0.9$. Interesting pattern, with coexisting high 
and low particle-density domains, is visible. A log-log plot of the decay of kinetic energy for the system, as a function of $t$, is presented 
in Figure \ref{fig6}(b). The initial decay (corresponding to HCS) is consistent with the prediction of Haff \cite{haff}, 
$E \sim e^{-a\tau}$ ($a$ is a constant; analytical 
curve is not shown). The late time data follow a power-law in $t$, with exponent $-1$. This is, thus, 
consistent with the prediction of Eq. (\ref{mass_t}). 
In Figure \ref{fig6}(c) we show a log-log plot of $m$ vs $t$. The data in the late time scaling regime are seen to obey a power law, 
the exponent being $\lesssim 2/3$. 
Here we mention that in a previous work \cite{paul}, via a finite-size scaling analysis, 
we had shown that the average domain length grows as $t^{\alpha}$ with 
$\alpha \simeq 1/3$. The conclusion from Figure \ref{fig6}(c) is thus in agreement with this earlier study. 
Nevertheless, given that for the GGM there exists possibility 
of continuous change of density within the domains, it is instructive to calculate the average mass \cite{paul}. 

\par
~~Since Eq. (\ref{mass_t}) predicts inverse relationship between mass and energy, 
the kinetics of GGM is different from ballistic aggregation, particularly 
when the exponents do not follow even a hyperscaling relation \cite{trizac1,trizac2} (see discussion below in the context of 
BAM). 
Matching of the exponent for energy decay (with Eq. (\ref{mass_t})) is 
accidental. In the rest of the subsection, therefore, we focus only on the BAM. 
This is by keeping the primary objective of studying aging during 
ballistic aggregation in mind. 
\par
\begin{figure}[htb]
\centering
\includegraphics*[width=0.46\textwidth]{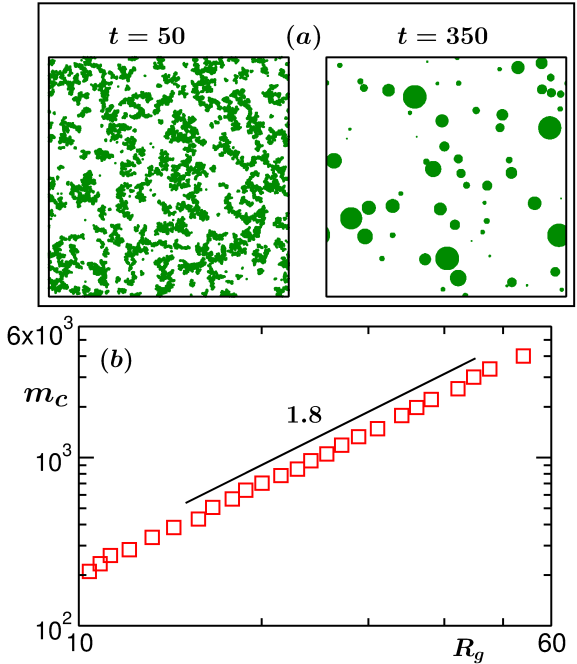}
\caption{\label{fig7} (a) (Left frame) A snapshot during the evolution of the fractal BAM in $d=2$ with $L=512$. 
See text for details. (Right frame) Same as the left frame but 
with spherical structural approximation and for $L=1024$. 
In both the frames only parts of the original systems are shown. Times are mentioned on top of the frames.
(b) Cluster mass, from a typical snapshot, is shown as a function of the radius of gyration, $R_g$,
for the fractal BAM case. The solid line there is a power-law with exponent $1.8$.
Rest of the results will be presented for $L=1024$ and spherical BAM.}
\end{figure}

~~ There are different variants of models dealing with
ballistic aggregation. E.g. there exists interest in a model where
ballistically moving particles from a source get deposited on a fixed substrate or seed. Such models are of relevance
in situations like construction of vapor-deposited thin films and the corresponding structures are fractal \cite{ramanlal}.
In the present case, however, all the clusters move ballistically, between collisions.
Simulation of such BAM in $d>1$ is not straight forward. 
If no deformation of the clusters is considered, highly 
fractal structures are expected in this situation as well. 
In that case, one needs to keep track of the exact points of contact, 
when two clusters collide. This is a difficult task, particularly if the 
rotations of the clusters are considered. In the left frame of Figure \ref{fig7}(a) 
we show a snapshot, obtained during an evolution for the BAM, without incorporating any deformation and 
considering only the translational motion of the clusters. Nice fractal pattern is seen. We
have estimated the fractal dimension which we discuss later.

\par
~~Because of the above mentioned difficulty in dealing with the actual physical scenario, 
researchers \cite{trizac1,trizac2,pathak} 
have adopted a spherical structural approximation. 
In this method, after a collision between two spherical objects of diameters 
$\sigma_1$ and $\sigma_2$, the mass of the resulting cluster is 
(usually uniformly) distributed over the volume of 
a sphere or circle (depending upon the system dimension) of diameter
\begin{equation}\label{ba_dia}
 \sigma = (\sigma_1^d + \sigma_2^d)^{1/d}.
\end{equation}
Many materials are prone to permanent deformation after high impact collisions. This is, thus, a reasonable 
approximation if the time scale of deformation is low, 
compared to the mean free time. In any case, given that fractality
offers larger collision cross-section, the dynamics of the systems with such spherical structural approximation
will be different from those without the approximation. In the rest of the subsection, 
unless otherwise mentioned, by BAM we will refer to the
ballistic aggregation model with circular approximation.

In the BAM, the post-collisional position and velocity of a new cluster can be obtained from
the conservation equations related to centre of mass and linear momentum.
A snapshot during the evolution of a system with such rules is shown in the right frame of Figure \ref{fig7}(a). 
Before presenting results on dynamics
of this simplified model, in Figure \ref{fig7}(b) we present result for the fractal dimension 
corresponding to the snapshot
in the left frame of Figure \ref{fig7} (a). Here we show mass of individual clusters as a function of
the radius of gyration ($R_g$), on a log-log scale. Nice power-law behavior is visible, providing
(mass) fractal dimension $d_f \simeq 1.8$.
As mentioned above, henceforth we will work with only the circular BAM. Even though the primary aim is to examine 
scaling property related to aging, in the following we present accurate results for energy decay and cluster growth as well, 
from appropriate analyses. To the best of our knowledge, such accurate analyses were not previously performed to draw 
conclusions on the behavior of these quantities.

\par
\begin{figure}[htb]
\centering
\includegraphics*[width=0.52\textwidth]{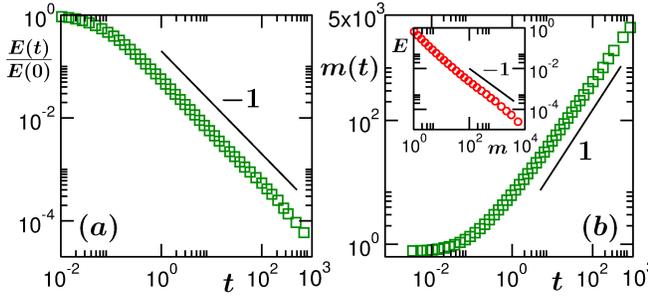}
\caption{\label{fig8} Log-log plots of the (a) kinetic energy vs time and (b) mass vs time, for the $d=2$ 
BAM. The solid lines in these 
figures are power-laws with exponents $-1$ and $1$, respectively. In the inset of (b) we show a log-log
plot of $E$ vs $m$. The solid there is a power-law with exponent $-1$.}
\end{figure}

~~In Figure \ref{fig8}(a) we show a log-log plot of energy decay as a function of time. 
A plot for the growth of mass is shown in Figure \ref{fig8}(b). 
Power laws 
in both the cases can be identified. While from these log-log plots it appears that the energy and mass are 
inversely proportional to each other, 
as predicted in Eq.(\ref{mass_t}), the inset of Figure \ref{fig8}(b), where we show
kinetic energy as a function of mass, provides a different information. There the exponent of the power-law
decay appears clearly higher than unity, approximately $1.15$, over a significant range.
For an accurate estimate we, thus, calculate the instantaneous exponents \cite{huse} for the time dependence of $m$ and $E$ as
\begin{equation}\label{instan_exp}
 \theta_i = -\frac{d ~{\rm{ln}} E}{d ~{\rm{ln}} t}, ~~~~~~\zeta_i = \frac{d ~{\rm{ln}} m}{d ~{\rm{ln}} t}.
\end{equation}
 Such exercises were performed for the $d=1$ BAM as well. However, we avoided presenting those results, 
since this aspect in $d=1$ is better understood.
\begin{figure}[htb]
\centering
\includegraphics*[width=0.52\textwidth]{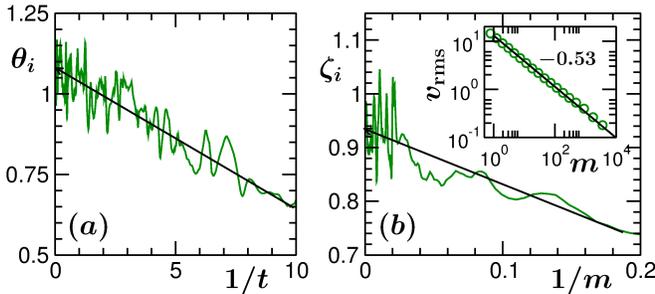}
\caption{\label{fig9} Plots of (a) $\theta_i$ vs $1/t$ and (b) $\zeta_i$ vs $1/m$, for the $2D$ 
BAM. The solid lines are guides to the eye.
Inset in (b): Log-log 
plot of $v_{\rm{rms}}$ vs $m$, for the $2D$ BAM.
 The solid line there is a power-law, exponent for which is mentioned next to it.}
\end{figure}

\par
~~We have plotted $\theta_i$, vs $1/t$, and $\zeta_i$, vs $1/m$, in Figures \ref{fig9} (a) and (b), respectively. 
In the asymptotic limit we obtain $\theta \simeq 1.08$ and $\zeta \simeq 0.94$. 
Thus, the predictions of Eq. (\ref{mass_t}) are not obeyed. 
These numbers, however, appear consistent with  a hyper-scaling relation \cite{trizac2} 
in $d=2$ (for ballistic aggregation):
\begin{equation}\label{hyper}
 \theta + \zeta =2.
\end{equation}
The failure of Eq. (\ref{mass_t}) lies in the fact that at 
low packing fraction the assumption related to uncorrelated velocity, inherent in the derivation of Eq. (\ref{mass_t}),
breaks down \cite{trizac3}. It is expected that at higher density, where the collision events are more frequent,
this prediction will work \cite{trizac2,trizac3,pathak}.
Here we ask the question is it not possible to obtain the above mentioned value of $\zeta$ from Eq. (\ref{kinetic}) or (\ref{frac_exponent})? 
Note that under the spherical approximation $d_f=2$. Thus, we need to estimate $\gamma$ to find out the reason for deviation of $\zeta$ from unity
(see Eq. (\ref{mass_t})).
\par
~~In the inset of Figure \ref{fig9}(b) we plot $v_{\rm{rms}}$ as a function of $m$. A power-law behavior from the log-log plot can be appreciated. 
The corresponding 
exponent ($\gamma \simeq 0.53$) provides $\zeta \simeq 0.97$ (see Eq. (\ref{frac_exponent})). 
Even though this number is smaller than $1$, no conclusive remark should be made from such small deviation.
Following Ref. \cite{trizac3}, we state here the reason behind a deviation between $\theta$ and
$\zeta$. Via the introduction of a dissipation parameter ($\alpha^{\prime}$), ratio between kinetic energy dissipation in 
a collision and mean kinetic energy per particle, these authors showed that $\alpha^{\prime}=1$ for high collision
frequency. On the other hand, for low frequency, i.e., at low particle
density, $\alpha^{\prime} > 1$. In the latter scenario, the particles with larger
kinetic energy than the mean undergo more frequent collisions, enhancing the dissipation. This leads to
a value of $\theta$ higher than unity. This fact becomes more prominent at densities smaller than 
the one considered here. E.g., for $\rho=0.005$, we find $\theta\simeq 1.15$ and $\zeta\simeq 0.85$.
Similar fact is observed in $d=3$. There, in future, we intend to verify how well the corresponding
hyperscaling relation \cite{trizac2} holds.
Next we present results for aging. 
\par
\begin{figure}[htb]
\centering
\includegraphics*[width=0.51\textwidth]{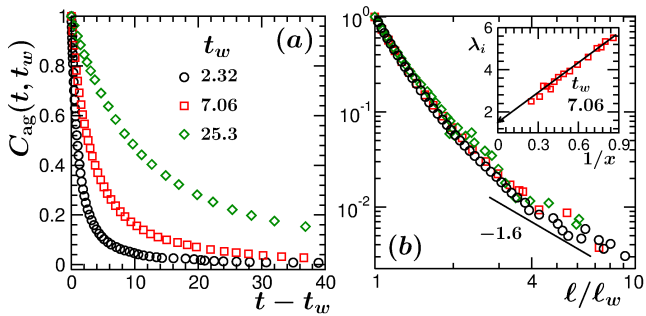}
\caption{\label{fig10} (a) For the $2D$ 
BAM the autocorrelation function $C_{\rm{ag}}(t,t_w)$ is plotted vs $t-t_w$. Results from 
three different $t_w$ values are included. (b) Log-log plot of $C_{\rm{ag}}(t,t_w)$ vs $x~(=\ell/\ell_w)$, using data sets of (a). 
The solid line represents a power-law 
with exponent $\lambda=1.6$. In the inset of (b) we show the instantaneous exponent $\lambda_i$ as a function of $1/x$. The solid line there 
is a linear extrapolation to $x=\infty$.}
\end{figure}

~~ In Figure \ref{fig10}(a) we show plots of $C_{\rm{ag}}(t,t_w)$, vs $t-t_w$, for a few different values of $t_w$. 
Like in $d=1$, time translational invariance is 
absent, as expected. It is clearly seen that with increasing age relaxation gets slower. In Figure \ref{fig10}(b) 
we show the log-log plots of $C_{\rm{ag}}(t,t_w)$ as 
a function of $\ell/\ell_w$. Very nice collapse of data on a master curve is visible. 
This confirms the scaling property of Eq. (\ref{aging1}). For large values of $\ell/\ell_w$
power-law decay becomes prominent. Continuous 
bending of the master curve for small abscissa variable implies early-time correction to the power-law. 
The large $x$ data appear to be consistent with an exponent $\lambda \simeq 1.6$, the 
number being roughly the same as in the $d=1$ case. For aging in kinetics of phase transitions, 
on the other hand, one observes strong dimension dependence of $\lambda$ \cite{mid2,mid3}. 
Here note that in 
$d=2$ we expect \cite{yeung2} $\beta=4$. Thus, the (lower) bound in (\ref{aging_bound}) is $3$. 
This calls for a look at the behavior of the equal-time structure factor for the 
present problem. While for bicontinuous domain structures ($d>1$) the analytical prediction ($\beta=4$) 
has been numerically confirmed \cite{yeung2,ahmad}, the cases of 
discrete domain morphology are less studied. Before taking a look at the equal time structure factor, since a violation of the 
bound appears to be a possibility, in the inset of Figure \ref{fig10}(b) we plot the instantaneous exponent 

\begin{equation}\label{instan_exp_aging}
 \lambda_i = - \frac{d ~{\rm{ln}} C_{\rm{ag}}}{d ~{\rm{ln}} x},
\end{equation}
as a function of $1/x$, to accurately quantify $\lambda$. The data set provides an asymptotic value $\lambda \simeq 1.55$.

\begin{figure}[htb]
\centering
\includegraphics*[width=0.39\textwidth]{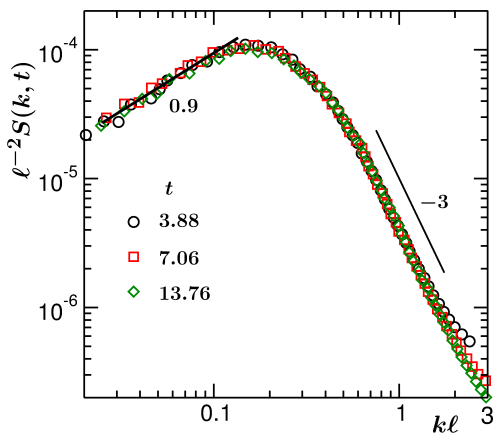}
\caption{\label{fig11} Log-log plot of $\ell^{-2}S(k,t)$ vs $k\ell$, for the BAM in $d=2$.
 The solid lines are power laws, exponents for which are mentioned 
on the figure.}
\end{figure}

\par
~~In Figure \ref{fig11} we present a scaling plot of $S(k,t)$, viz., we show $\ell^{-2}S(k,t)$ vs $k\ell$, on a log-log scale. 
The large $k$ data are consistent with the 
power-law exponent $-3$, that corresponds to the Porod law \cite{porod} in $d=2$ for a scalar order parameter. 
In the small $k$ region, on the other hand, the 
enhancement is much weaker than $k^4$. In fact, in the relevant region, we observe $\beta\lesssim 1$. 
Similar number we have observed in recent studies of kinetics of phase 
transition with conserved off-critical composition as well, for which one naturally 
obtains circular or spherical domain structures.
 In that case we have the lower  bound 
to be $\lesssim 1.5$, which is satisfied by the above estimated value of $\lambda$.

\section{conclusion}
~~ We have studied the kinetics of clustering in models of granular gas (GGM)
and ballistic aggregation (BAM), in $d=1$ and $2$. 
It is shown that 
the average size of the clusters grows as power-law with time.
In $d=1$, via a dynamic renormalization group theoretical method of analysis \cite{rol}, the exponent 
for the GGM has
been identified to be approximately $2/3$, in agreement with that for the BAM. 
In this dimension, for GGM as well as BAM, 
the growth appears inversely proportional to the energy decay, showing consistency with 
the scaling predictions of Carnevale {\it{et al.}} \cite{car} for ballistic aggregation. 
The growth mechanism, for the GGM case, has been identified directly by calculating 
the mean-squared-displacements \cite{han}
of the centres of mass of clusters before they undergo collisions. To avoid the
inelastic collapse, for the GGM, in this dimension, we have used a cut-off $\delta$. For
relative velocities $<\delta$, the value of $e$ was set to unity for the colliding
partners \cite{ben}. We observed $\delta$-dependent saturation in the growth of mass, 
appearing earlier for larger values
of $\delta$. Interestingly, in such saturation regime also the energy decay continued to follow
the theoretical scaling form $t^{-2/3}$. This calls for further investigation. In $d=2$, on the other hand, 
any equivalence between GGM and BAM is shown to be absent. 
\par
~~In both the dimensions, for the density field, we have studied the aging property \cite{puri,fish} for ballistic 
aggregation, which is first in the literature.
It is shown that, 
like in kinetics of phase transitions, the order-parameter autocorrelation function scales with 
$\ell/\ell_{w}$. The asymptotic forms of the scaling functions have been identified to be  power-laws. 
The corresponding exponents have been estimated and 
discussed with reference to the structural property. It is shown that the exponents obey dimension dependent 
lower bounds, \cite{yeu} predicted for kinetics of phase transitions where systems move towards 
a new equilibrium. However, unlike in the kinetics of phase transitions, the aging exponents 
here appear to be very close to the lower bounds, irrespective of the dimension.
The similar values of the exponent for the GGM and
the BAM cases in $d=1$, further suggests close equivalence between the dynamics in the
two cases, in this dimension. We intend to undertake similar studies in $d=3$.
\par
~~With respect to the more realistic ballistic aggregation, 
the simulations are rather challenging for $d>1$. This is because
of the formation of fractal structures. Because of this reason, like in the existing
simulation studies, spherical structural approximation has been used by us.
It will be interesting to investigate the scaling properties related to aging 
and other aspects without such approximation.

~~{\bf Acknowledgment:} The authors thank Department of Science and Technology, India, for financial support.
SKD also acknowledges the Marie Curie Actions Plan of European Commision 
(FP7-PEOPLE-2013-IRSES grant No. 612707, DIONICOS) as well as International Centre for
Theoretical Physics, Trieste, for partial supports.
SP is thankful to UGC, India, for research fellowship.

~${*}$ das@jncasr.ac.in


\end{document}